\documentclass[review]{elsarticle}

\usepackage{lineno,hyperref}
\modulolinenumbers[5]

\journal{Journal of \LaTeX\ Templates}

\usepackage{amsthm}
\usepackage{amsmath}
\usepackage{amssymb}
\usepackage{graphicx}
\usepackage{xspace}

\usepackage{url}
\usepackage{algorithmic}
\usepackage{algorithm} 
\algsetup{linenosize=\scriptsize}
\xspaceaddexceptions{=}
\xspaceaddexceptions{\}}
\xspaceaddexceptions{\in}
\algsetup{linenosize=\scriptsize}
\usepackage{algorithm} 
\usepackage[bf]{subfigure}
\usepackage[pdftex,table,fixpdftex]{xcolor}
\usepackage{longtable}
\usepackage{threeparttable}
\usepackage{multirow}
\usepackage{tabularx}
\usepackage{hhline}

\newcommand{\eg}{e.g.,\xspace}

\newcommand{\agent}{\ensuremath{i}\xspace}
\newcommand{\child}{\ensuremath{u}\xspace}
\newcommand{\plan}{\ensuremath{j}\xspace}

\newcommand{\timePoint}{\ensuremath{t}\xspace}

\newcommand{\numOfAgents}{\ensuremath{n}\xspace}
\newcommand{\numOfPlans}{\ensuremath{v}\xspace}
\newcommand{\numOfChildren}{\ensuremath{k}\xspace}
\newcommand{\horizon}{\ensuremath{T}\xspace}

\newcommand{\signal}[1]{\ensuremath{S_{#1}}\xspace} 
\newcommand{\signalValue}[2]{\ensuremath{s_{#1,#2}}\xspace} 

\newcommand{\shuffledSignal}[1]{\ensuremath{\hat{S}_{#1}}\xspace} 
\newcommand{\shuffledSignalValue}[2]{\ensuremath{\hat{s}_{#1,#2}}\xspace} 

\newcommand{\windowIndex}{\ensuremath{w}\xspace}

\newcommand{\numOfWindows}[1]{\ensuremath{q_{#1}}\xspace} 
\newcommand{\windows}[1]{\ensuremath{W_{#1}}\xspace} 
\newcommand{\window}[2]{\ensuremath{S_{#1,#2}}\xspace} 
\newcommand{\windowValue}[3]{\ensuremath{s_{#1,#2,#3}}\xspace} 

\newcommand{\slotIndex}{\ensuremath{o}\xspace}
\newcommand{\numOfSlots}[1]{\ensuremath{K_{#1}}\xspace} 
\newcommand{\numOfWindowSlots}[2]{\ensuremath{k_{#1,#2}}\xspace} 
\newcommand{\slots}[2]{\ensuremath{Q_{#1,#2}}\xspace} 
\newcommand{\rankedSlots}[2]{\ensuremath{\hat{Q}_{#1,#2}}\xspace} 
\newcommand{\slot}[3]{\ensuremath{S_{#1,#2,#3}}\xspace} 
\newcommand{\slotValue}[4]{\ensuremath{s_{#1,#2,#3,#4}}\xspace} 
\newcommand{\slotSize}[2]{\ensuremath{l_{#1,#2}}\xspace} 

\newcommand{\EVModel}{\ensuremath{\mathsf{m}}\xspace}
\newcommand{\chargeRate}[1]{\ensuremath{r_{\mathsf{#1}}}\xspace} 
\newcommand{\batteryCapacity}[1]{\ensuremath{b_{#1}}\xspace} 
\newcommand{\minChargingTime}{\ensuremath{m}\xspace}
\newcommand{\chargeTime}[2]{\ensuremath{T_{\mathsf{c}}(#1,#2)}\xspace} 
\newcommand{\chargeIntervals}[2]{\ensuremath{z_{\mathsf{i}}(#1,#2)}\xspace}

\newcommand{\EVUsage}[1]{\ensuremath{U_{#1}}\xspace} 
\newcommand{\EVUsageValue}[2]{\ensuremath{u_{#1,#2}}\xspace} 
\newcommand{\discomfort}[1]{\ensuremath{g_{#1}}\xspace} 
\newcommand{\totalDiscomfort}{\ensuremath{G_{\mathsf{d}}}\xspace} 
\newcommand{\fairness}{\ensuremath{G_{\mathsf{f}}}\xspace} 

\newcommand{\price}{\ensuremath{P}\xspace}
\newcommand{\priceValue}[1]{\ensuremath{p_{#1}}\xspace} 

\newcommand{\power}[1]{\ensuremath{e_{#1}}\xspace} 

\newcommand{\selectionFunctionIndex}{\ensuremath{\mathsf{s}}\xspace} 
\newcommand{\selectionFunction}[2]{\ensuremath{f_{#1}(#2)}\xspace} 

\newcommand{\demand}[3]{\ensuremath{d_{#1,#2,#3}}\xspace} 
\newcommand{\aggregateDemand}[2]{\ensuremath{a_{#1,#2}}\xspace} 
\newcommand{\demandPlan}[2]{\ensuremath{d_{#1,#2}}\xspace}
\newcommand{\oldDemandPlan}[1]{\ensuremath{\hat{d}_{#1}}\xspace}
\newcommand{\demandPlans}[1]{\ensuremath{D_{#1}}\xspace}
\newcommand{\aggregatePlan}[1]{\ensuremath{a_{#1}}\xspace}
\newcommand{\oldAggregatePlan}[1]{\ensuremath{\hat{a}_{#1}}\xspace}

\newcommand{\minCost}{\textsc{\scriptsize MIN-COST}\xspace}
\newcommand{\minDev}{\textsc{\scriptsize MIN-DEV}\xspace}

\DeclareMathOperator*{\argmin}{arg\,min}

\begin{document}
\linespread{1}

\begin{frontmatter}

\title{Socio-technical Smart Grid Optimization via Decentralized Charge Control
of Electric Vehicles}

\author{Evangelos Pournaras\fnref{myfootnote}}
\fntext[myfootnote]{Stampfenbachstrasse 48, 8092, Zurich, Switzerland, Email: \{epournaras,ysrivatsan,zhuiting,fangx\}@ethz.ch}
\author{Seoho Jung\fnref{seohofootnote}}
\fntext[seohofootnote]{Chair of Micro and Nanosystems, Tannenstrasse 3
8092, Zurich, Switzerland, Email: seoho.jung@micro.mavt.ethz.ch}
\author{Srivatsan Yadhunathan\fnref{myfootnote}}
\author{Huiting Zhang\fnref{myfootnote}}
\author{Xingliang Fang\fnref{myfootnote}}

\address{ETH Zurich, Zurich, Switzerland}

\begin{abstract}
The penetration of electric vehicles becomes a catalyst for the sustainability of Smart Cities. However, unregulated battery charging remains a challenge causing high energy costs, power peaks or even blackouts. This paper studies this challenge from a socio-technical perspective: social dynamics such as the participation in demand-response programs, the discomfort experienced by alternative suggested vehicle usage times and even the fairness in terms of how equally discomfort is experienced among the population are highly intertwined with Smart Grid reliability. To address challenges of such a socio-technical nature, this paper introduces a fully decentralized and participatory learning mechanism for privacy-preserving coordinated charging control of electric vehicles that regulates three Smart Grid socio-technical aspects: (i) reliability, (ii) discomfort and (iii) fairness. In contrast to related work, a novel autonomous software agent exclusively uses local knowledge to generate energy demand plans for its vehicle that encode different battery charging regimes. Agents interact to learn and make collective decisions of which plan to execute so that power peaks and energy cost are reduced system-wide. Evaluation with real-world data confirms the improvement of drivers' comfort and fairness using the proposed planning method, while this improvement is assessed in terms of reliability and cost reduction under a varying number of participating vehicles. These findings have a significant relevance and impact for power utilities and system operator on designing more reliable and socially responsible Smart Grids with high penetration of electric vehicles.
\end{abstract}


\end{frontmatter}


\section{Introduction}

The growing scale and complexity of urban environments as well as the implications of climate change on the planet make the use of renewable energy resources and the elimination of fossil fuels imperative for meeting sustainable development goals~\cite{Griggs2013}. A progressive step towards this direction is the large adoption of electric vehicles (EVs) to replace internal combustion engines. Instead of vehicles that run on gasoline and diesel, electric vehicles are equipped with batteries and they are charged when plugged into the power grid. Once charged, vehicles can travel up to a few hundred kilometers before requiring a recharge. Currently, private electric vehicles are predominantly charged at home, but more public charging stations are established. It is predicted that by 2025, electric vehicles will share approximately 22\% of the current vehicle market~\cite{MarketProjectionGS}. However, without a thorough investigation of the potential impact that a large-scale adoption of electric vehicles can have on existing power infrastructure, a relatively sudden influx in power demand can potentially have disastrous consequences on the power grid reliability.

The charging of an electric vehicle is usually performed soon after its driver returns home or arrives at work and plugs the vehicle to the grid. Given that a large number of drivers return home or arrive at work around the same time of the day, the synchronized charging of vehicles can result in a power peak demand~\cite{tan2016integration}. Moreover, the development of battery technologies often aims at reducing charging times~\cite{shen2014optimization}. This makes power peaks as well as power influxes sharper and, as a result, costly to serve and manage. From the supply-side, the integration of renewable energy resources makes power supply more volatile and complex to plan so that they match the power demand of electric vehicles given the stochastic nature of renewables~\cite{kempton2005vehicle}. Power utilities have the option to control the charging of electric vehicles to improve power grid reliability~\cite{tan2016integration, green2011impact}, which however requires a computational infrastructure that may be costly and unscalable to centrally manage, i.e. single point of failure. Moreover, centralized charging control requires the acquisition of sensitive personal data such as mobility patterns and residential occupancy information that can violate privacy and allow discriminatory data analytics, which can in turn undermine the adoption of electric vehicles and the trust of citizens on this technology~\cite{green2011impact,Christin2011,Han2016}. Finally, charging control is not exclusively a technical challenge, dealt mainly as such in earlier work (see below), but a social one as well. Comfort and fairness criteria in terms of convenient charging times for the citizens and equal convenience for all citizens can make electric vehicles a more viable adopted technology. 

This paper introduces a novel \emph{Vehicle to Grid}~\cite{tan2016integration} (V2G) socio-technical control paradigm of coordinated charging applied as a Smart Grid enabler. Electric vehicles are not required to return electricity to the grid but instead they provide demand-response services to power utilities using their coordination capabilities. Therefore, this proposed novel methodology neither violates physical constraints nor operationally depends on characteristics of the power infrastructure. Moreover, in terms of reactive power and voltage control, earlier work provides an optimal scheduling formulation without violating the grid operating constraints~\cite{Wang2019}. Coordination is distinguished from a localized distributed optimization as the former one requires interactions and the exchange of information between electric vehicles such that a system-wide objective is satisfied. For instance, load-balancing or matching of supply-demand, i.e. minimization of variance, root mean square error or other complex quadratic cost functions~\cite{Rockafellar2000}, are examples of such objectives. When charging control of electric vehicles is performed by autonomous (software) agents acting on behalf of their drivers and agents locally generate multiple (alternative) charging plans as contributions to the operational flexibility of the Smart Grid, the charging coordination problem turns out to be a 0-1 multiple-choice combinatorial optimization problem, which is NP-hard. A fully decentralized and privacy-preserving learning algorithm is employed to tackle this challenging computational problem: I-EPOS (\url{http://epos-net.org}), the \emph{Iterative Economic Planning and Optimized Selections}~\cite{Pilgerstorfer2017,Pournaras2018}. Socio-technical aspects are studied such as the impact of coordination on human discomfort and fairness measured by the likelihood of disturbing the usual mobility and lifestyle habits of citizens when alternative charging regimes are adopted. This perspective makes the proposed approach more realistic, viable and applicable compared to related methodologies reviewed below.

Related work on charging control of electric vehicles is classified in three categories: (i) Non-coordinated optimization of battery utilization to improve environmental factors and the individual driving profile of a vehicle~\cite{Hu2016,Moura2011} as well as the multi-objective optimization of plug-in hybrid electric vehicles~\cite{Du2018,Chen2018,Ahmadi2019}, e.g. fuel economy and motor start engine. These approaches do not focus on the optimization and reliability of the Smart Grid. (ii) Centralized coordination of charging control via single aggregators that provide ancillary services, a category in which 47 reviewed methodologies fall in~\cite{Bessa2012,Deilami2011,bahrami2014game,chung2016ev,Wenzel2018,Shamsdin2019,Chen2019,Wang2019}. (iii) Distributed methodologies that focus on localized cost minimization~\cite{Ma2013,Valogianni2014,wen2012decentralized,Gan2013} rather than decentralized charging coordination of autonomous electric vehicles that is the focus of this paper (Novelty 1).

In the third category falls a technical study on regulating a fair access of the vehicles to charging, i.e. proportional, max-min, minimum delay and proportional utility~\cite{Ardakanian2013}. Similarly, notions of convenience are earlier introduced and measured by the \emph{state of charge} (SoC) and the total charging time of the battery~\cite{chung2016ev,wen2012decentralized}. The driver's satisfaction in terms of charging at the shortest time is studied in three centralized charging policies~\cite{Shamsdin2019}. None of these methodologies tackle social aspects of Smart Grid optimization such as discomfort and fairness measured empirically by the driving behavior and human mobility. Social discomfort and fairness are so far studied in the context of residential energy and appliances~\cite{Nguyen2014,Pournaras2014,pournaras2014decentralized}, e.g. heating and cooling. Recent state of the art studies notions of discomfort, i.e. changes of regular travel times or long term battery degradation, as a result of a centralized optimization of shifted charging times~\cite{Xu2018,Kikusato2018}. In contrast, this paper advances state of the art by studying both discomfort and fairness in the context of a decentralized and autonomous charging control mechanism that is privacy-preserving by design (Novelty 2).

Finally, this paper measures the impact of a varying participation level on the optimization process of charging electric vehicles. To the authors' best knowledge, this is not addressed in earlier work (Novelty 3). 

In summary, the literature gap that this paper fills is the following: \emph{how to improve the Smart Grid reliability in the scenario of decentralized and coordinated charging control of electric vehicles, while meeting social welfare criteria related to drivers' discomfort and fairness in the experienced discomfort}. The main contributions of this paper are the following:

\begin{itemize}
\item A new privacy-preserving methodology to locally and autonomously generate charging plans for electric vehicles by reasoning based on local historical mobility data and technical vehicle characteristics.
\item The applicability of I-EPOS, a general-purpose decentralized learning algorithm, to the charging control of electric vehicles for the socio-technical optimization of Smart Grids. 
\item Quantitative findings using real-world data that show how the optimization of technical objectives, e.g. reliability, influence the user discomfort and social fairness. 
\item Quantitative findings that show how all socio-te\-chni\-cal aspects are influenced by a varying participation level of electric vehicles in the optimization process. 
\item A new application scenario and benchmark dataset~\cite{Pournaras2019b} for the evaluation of combinatorial optimization algorithms. 
\end{itemize}

\begin{figure}[!htb]
\centering
\includegraphics[width=0.8\textwidth]{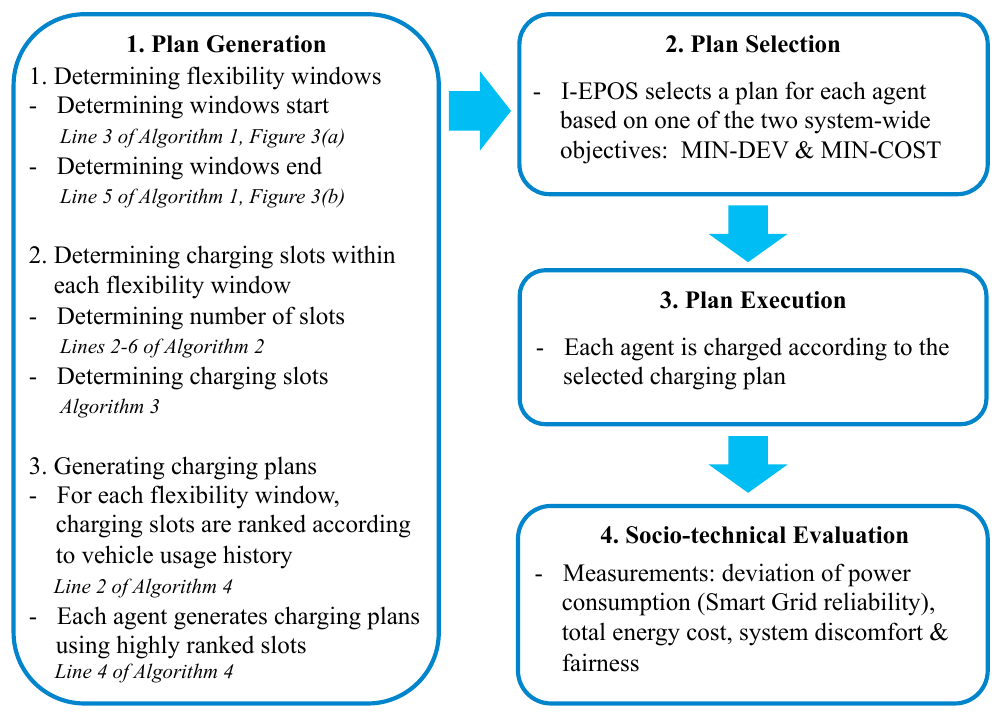}
\caption{A schematic overview of the contributed solution.}\label{fig:flow-chart}
\end{figure}

Figure~\ref{fig:flow-chart} provides a schematic reference of the solution. This paper is outlined as follows: The following section introduces the local operational planning for electric vehicles. Section~\ref{sec:decision-making} illustrates the decentralized collective learning process as well as it defines measures of reliability, discomfort and fairness employed for the socio-technical Smart Grid optimization. Section~\ref{sec:evaluation} illustrates how planning is performed using a real-world dataset and illustrates the experimental results and findings. Finally, Section~\ref{sec:conclusion} concludes this paper and outlines future work. 

\section{Local and Autonomous Operational Planning}\label{sec:model}

This section illustrates how electric vehicles can locally and autonomously plan, i.e. schedule, their power usage in order to achieve various system-wide objectives such as the improvement of system reliability or the reduction of power costs. Table~\ref{table:math-symbols} outlines the mathematical symbols used in this paper in the order they appear.

\begin{table}[!htb]
\footnotesize\centering
\caption{An overview of the mathematical symbols.}
\begin{tabularx}{\textwidth}{lX}
\hline
Symbol & Interpretation \\ \hline
\numOfAgents & Number of agents \\
$\agent \in \{1,...,\numOfAgents\}$ & An agent \\ 
\EVModel & Electric vehicle model \\
\minChargingTime & Minimum charging interval \\
\numOfPlans & Maximum number of generated plans \\
$\plan \in \{1,...,\numOfPlans\}$ & A generated plan \\
\demandPlans{\agent} & Generated plans of agent \agent \\
\demandPlan{\agent}{\plan} & Plan \plan of agent \agent \\
$\horizon$ & Planning horizon \\
\selectionFunction{\selectionFunctionIndex}{\demandPlans{\agent}} & Selection function \selectionFunctionIndex of a plan for execution \\ 
$\timePoint \in \{1,...,\horizon\}$ & Time point \\
\price & Price signal \\
\priceValue{\timePoint} & Price value at time \timePoint \\
\signal{\agent} & State of charge signal of agent \agent \\
\signalValue{\agent}{\timePoint} & State of charge of agent \agent at time \timePoint \\
$\windowIndex \in \{1,...,\numOfWindows{\agent}\}$ & A window \\
\windows{\agent} & Windows of agent \agent \\
\window{\agent}{\windowIndex} & Window \windowIndex of agent \agent \\
\numOfWindows{\agent} & Number of windows of agent \agent \\
\windowValue{\agent}{x}{\windowIndex} & State of charge of agent \agent at the beginning $x$ of window \windowIndex \\
\slotSize{\agent}{\windowIndex} & Slots size of agent \agent in window \windowIndex \\
\chargeTime{\EVModel}{\windowValue{\agent}{x}{\windowIndex}} & Charging time of model \EVModel with state of charge \windowValue{\agent}{x}{\windowIndex} at the beginning $x$ of window \windowIndex \\
\batteryCapacity{\EVModel} & Battery capacity of model \EVModel \\
\chargeRate{\EVModel} & Charging rate of model \EVModel \\
$\slotIndex \in \{1,...,\numOfWindowSlots{\agent}{\windowIndex}\}$ & A slot \\
\numOfWindowSlots{\agent}{\windowIndex} & Number of slots of agent \agent in window \windowIndex \\
\slot{\agent}{\windowIndex}{\slotIndex} & Slot \slotIndex of agent \agent in window \windowIndex \\
\slots{\agent}{\windowIndex} & Slots of agent \agent in window \windowIndex \\
\numOfSlots{\agent} & Numbers of slots for agent \agent \\
\slotValue{\agent}{\timePoint}{\windowIndex}{\slotIndex} & State of charge of agent \agent at time \timePoint of slot \slotIndex in window \windowIndex \\
\EVUsage{\agent} & Likelihood of usage for vehicle \agent \\
\EVUsageValue{\agent}{\timePoint} & Likelihood of usage at time \timePoint for vehicle \agent \\
\discomfort{\agent} & Discomfort of a plan generated by agent \agent \\
\demand{\agent}{\plan}{\timePoint} & Power demand of agent \agent at time \timePoint for plan \plan \\
\chargeIntervals{\EVModel}{\windowValue{\agent}{x}{\windowIndex}} & Number of charging intervals for model \EVModel with state of charge \windowValue{\agent}{x}{\windowIndex} at the beginning $x$ of window \windowIndex \\
\rankedSlots{\agent}{\windowIndex} & Ranked slots of agent \agent in window \windowIndex \\
\shuffledSignal{\agent} & Shuffled state of charge signal of agent \agent \\
\shuffledSignalValue{\agent}{\timePoint} & State of charge from a shuffled signal of agent \agent at time \timePoint\\
\power{\EVModel} & Power demand of model \EVModel when charging \\
$\sigma()$ & Standard deviation of a plan \\
\oldAggregatePlan{\agent} & Earlier aggregate plan of agent \agent \\
\oldDemandPlan{\agent} & Earlier selected plan of agent \agent \\
\aggregatePlan{\agent} & Current aggregate plan of agent \agent \\
\child & The child of an I-EPOS agent \\
\aggregateDemand{\agent}{\timePoint} & The power demand in the aggregate plan of agent \agent at time \timePoint \\
\totalDiscomfort & The average discomfort of all agents\\
\fairness & The fairness among the discomfort of all agents \\
$E_\mathtt{t}$ & The energy of a vehicle trip\\
$d_\mathtt{t}$ & The distance of a vehicle trip\\
$\eta$ & Energy conversion between gasoline and Joules\\
$f_e$ & Efficiency of an electric vehicle in city/highway\\
$\nu$ & Speed of a vehicle trip\\
$t_\mathtt{t}$ & Time of a vehicle trip\\
\hline
\end{tabularx}\label{table:math-symbols}
\end{table}

\subsection{An overview}\label{sec:model-overview}

This paper introduces a new concept of local and autonomous operational planning of electric vehicles as the means to meet system-wide objectives of Smart Grids. The motivation here is that if adjustments in power demand can be pre-computed and scheduled, operational uncertainties are minimized and more effective regulatory actions can be applied under several operational scenarios, \eg failures of power generators, price peaks, weather events influencing the availability of renewable energy resources, etc. Planning is a well-established approach in literature~\cite{Georgeff1988,DeWeerdt2005,Konolige1980,Maliah2017} and in several related real-world application domains~\cite{Pilgerstorfer2017,Pournaras2018,Barbati2012,Pournaras2017}.

Each vehicle is equipped with a software agent that can autonomously generate a number of \emph{possible charging plans} that determine how the vehicle is charged and draws power from the Smart Grid. Each plan may cause a varied level of driver's discomfort measured empirically via the likelihood of the plan to be violated by, for example, an unexpected traveling event. Agents make coordinated selections of a plan to execute such that they collectively accomplish a system-wide Smart Grid objective, while driver's discomfort is measured and self-regulated.

\begin{figure*}[!htb]
\centering
\includegraphics[width=1.0\textwidth]{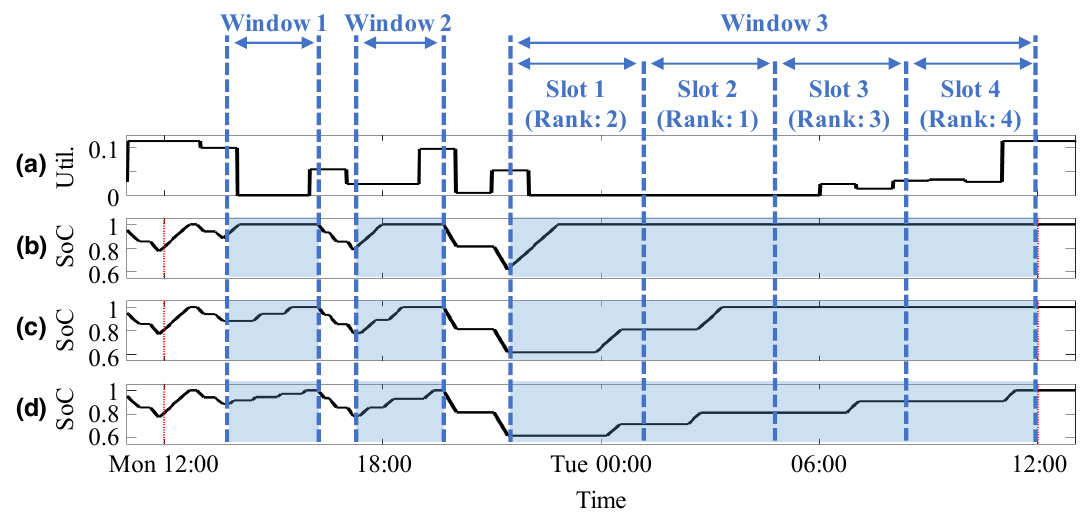}
\caption{Plan generation defines flexibility windows within which the state of charge (SoC) increases or does not vary. In each window, several charging slots are defined in which a full charging can be performed. Slots are ranked from low to high according to the likelihood of vehicle usage computed from historical data. (a). Likelihood of the vehicle usage derived from historical data. (b). Original states of charge if the vehicle is charged at the beginning of each window. (c). States of charge in a generated plan that uses two slots with the lowest likelihood of usage in each window (e.g. Slot 2 and Slot 1 in Window 3). (d). States of charge in a generated plan that uses all slots in each window.}\label{fig:generation}
\end{figure*}

Figure~\ref{fig:generation} illustrates the proposed novel concept of plan generation that is the core contribution of this paper. A\-ge\-nts locally calculate the time required for a full battery charge as well as compute \emph{flexibility windows} within which their vehicle is usually available for charging. In practice, windows correspond to times in which a vehicle is parked at home. During these times, the state of charge of the battery in a vehicle increases, i.e. the vehicle is charged, or remains at the same level. The state of charge cannot decrease as the vehicle is parked and does not consume power from its battery. In each window, \emph{charging slots} can be efficiently computed. In each charging slot a full battery charging can be performed. A single charging slot up to a maximum number of slots that is equal to the maximum number of possible plans \numOfPlans form a flexibility window. The agent ranks the slots from low to high by a data-driven reasoning based on historical data: the likelihood of vehicle usage computed from past trips determines the ranking of the slots. Each charging plan uses a varied number of top ranked slots. For example, in a plan that uses two charging slots, the charging is completed in two different periods corresponding to the top-two ranked slots.

\begin{figure}[!htb]
\centering
\includegraphics[width=0.3675\columnwidth]{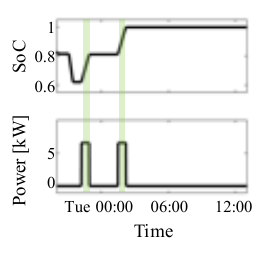}
\includegraphics[width=0.3675\columnwidth]{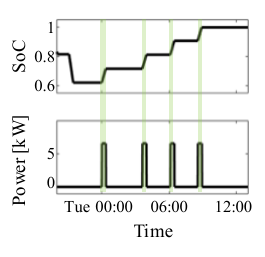}
\caption{State of charge (SoC) and power consumption of the Nissan model for Window 3 belonging to the plans of Figure~\ref{fig:generation}c and~\ref{fig:generation}d}\label{fig:plans}
\end{figure}

Figure~\ref{fig:plans} illustrates the redistributed power usage of the plans in Figure~\ref{fig:generation}c and~\ref{fig:generation}d for Window 3. Each electric vehicle model consumes a certain level of power from the grid when it is charged. For this illustration, a Nissan Leaf model equipped with a 6.6 kW onboard charger is used~\cite{NissanPower}.

Note that all planning operations are localized to preserve privacy that is a novelty of this approach. No personal data are shared to third parties. Technology for the realization of such electric vehicle charging control and planning exists~\cite{Crombez2013,Hayashi2015,Fincham2017,Hoque2017,Jiang2018} and can be adopted in the context of a (i) home energy management system~\cite{Hu2013}, (ii) a demand-response program by a power utility company~\cite{Rassaei2015} or (iii) a third party technology provider for electric vehicles~\cite{Pryor2008}.

\subsection{Technical concept}\label{sec:model-concept}


This section elaborates on the proposed technical concept. Assume a number of \numOfAgents electric vehicles powered by a common power system. Each vehicle is characterized by a model \EVModel that determines features such as the battery capacity and charging rate. Production models of electric vehicles are earlier reviewed~\cite{Sierzchula2012}. Each vehicle is assumed equipped with an agent \agent that is a software application controlling the battery charging. As earlier discussed, technology for such control is available in the market~\cite{Lacroix2011}. The agent is constrained by a minimum time \emph{interval} of size \minChargingTime during which the vehicle continuously charges without pause. The minimum time interval \minChargingTime constraints the number of interruptions on battery charging such that the battery lifetime is not heavily influenced. Moreover, it limits the computational search space, i.e. the larger the interval \minChargingTime the fewer the number of time shifts that can be performed on the battery charging within a certain period of time.  

Each agent \agent locally generates a sequence of a maximum of \numOfPlans \emph{plans} $\demandPlans{\agent}=(\demandPlan{\agent}{\plan})_{\plan=1}^{\numOfPlans}$ that schedule the power consumption of the vehicle for the future period $\horizon=|\demandPlan{\agent}{\plan}|$ when charging from the Smart Grid. These plans may be equivalent for the driver of the vehicle or they may cause different levels of \emph{discomfort}, for example, each plan may disturb the regular use of a vehicle to a different extent. Each agent \agent selects one and only one plan $\demandPlan{\agent}{\plan}=(\demand{\agent}{\plan}{\timePoint})_{\timePoint=1}^{\horizon}$ to execute according to a selection function $\plan=\selectionFunction{\selectionFunctionIndex}{\demandPlans{\agent}} \in \{1,...,\numOfPlans\}$ designed to serve a system-wide objective such as the improvement of reliability or the reduction of power cost in the Smart Grid. \emph{Reliability} concerns how homogeneous the total power demand is over time so that power peaks causing cascading failures~\cite{pournaras2014decentralized,Thapa2017} are prevented or mitigated. \emph{Cost} concerns the monetary value of the total power demand governed by the spot price signal $\price=(\priceValue{\timePoint})_{\timePoint=1}^{\horizon}$ (USD/kWh) of a power market~\cite{spotPrice}. 

Reasoning about plan generation is locally performed based on accumulated historical data that represent a typical temporal usage pattern of the electric vehicle and the driver's profile, e.g. daily or weekly usage. These data are not shared to any third party and managed by the agent to preserve privacy. They are referred to as $\signal{\agent}=(\signalValue{\agent}{\timePoint})_{\timePoint=1}^{\horizon}$, where $\signalValue{\agent}{\timePoint} \in [0,1]$ stands for the state of charge of the vehicle at time $\timePoint$. This signal is used as a seed for the plan construction using real-world data as illustrated in Section~\ref{subsec:methodology}. The goal of plan generation is to compute several ways of charging the electric vehicle during times in which the vehicle is not used, \eg when the vehicle is parked at home. These times are referred to in this paper as the \emph{flexibility windows}, $\windows{\agent}=(\window{\agent}{\windowIndex})_{\windowIndex=1}^{\numOfWindows{\agent}} \subseteq \signal{\agent}$, of an agent \agent. Algorithm~\ref{alg:windows} illustrates how windows are computed. 

\begin{algorithm}[!htb]
\centering
\small{
\begin{algorithmic}[1]
\REQUIRE \signal{\agent}
\STATE \windowIndex=0;
\FOR{$\timePoint=1$ to \horizon}
\IF{$\signalValue{\agent}{\timePoint}<\signalValue{\agent}{\timePoint-1}$ \AND $\signalValue{\agent}{\timePoint}<\signalValue{\agent}{\timePoint+1}$}
\STATE $\windowIndex=\windowIndex+1$;
\WHILE{$\signalValue{\agent}{\timePoint}\geq\signalValue{\agent}{\timePoint-1}$ \AND $\signalValue{\agent}{\timePoint}\leq\signalValue{\agent}{\timePoint+1}$ \AND $\timePoint < \horizon$}
\STATE  $\window{\agent}{\windowIndex}=\window{\agent}{\windowIndex} \cup \signalValue{\agent}{\timePoint}$
\STATE  $\timePoint=\timePoint+1$;
\ENDWHILE
\IF{$|\window{\agent}{\windowIndex}|\geq\chargeTime{\EVModel}{\windowValue{\agent}{x}{\windowIndex}}$}
\STATE $\windows{\agent} = \windows{\agent} \cup \window{\agent}{\windowIndex}$ 
\ELSE
\STATE $\windowIndex=\windowIndex-1$;
\ENDIF
\ENDIF
\ENDFOR
\ENSURE \windows{\agent}, $\forall \agent \in \{1,...,\numOfAgents\}$ 
\end{algorithmic}
}
\caption{Computation of flexibility windows.}\label{alg:windows} 
\end{algorithm}

\begin{figure}[!htb]
\centering
\subfigure[Start of window]{\includegraphics[width=0.3675\columnwidth]{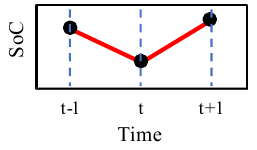}}
\subfigure[End of window]{\includegraphics[width=0.3675\columnwidth]{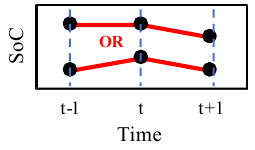}}
\caption{Detection of window limits. (a) State of charge is higher at $\timePoint+1$. (b) State of charge is equal or higher at $\timePoint-1$, and lower at $\timePoint+1$.}\label{fig:windows} 
\end{figure}

The algorithm identifies the times at which the state of charge stops decreasing and starts increasing (line 3 of Algorithm~\ref{alg:windows}). These times are the beginning of the flexibility windows. The end of the windows is detected by the times in which the state of charge stops increasing and starts decreasing (line 5 of Algorithm~\ref{alg:windows}). This indicates that the electric vehicle is again in use. Figure~\ref{fig:windows} illustrates how the start and end of a window are detected. The algorithm excludes windows that do not have the length for a full charging\footnote{The algorithm assumes that windows of length shorter than the full charging time are more sensitive to user interruptions and therefore have a higher uncertainty when used for scheduling the charging of the vehicle. The evaluation of this assumption with several other algorithm variations is subject of future work.} (line 9-13 of Algorithm~\ref{alg:windows}). The charging time \chargeTime{\EVModel}{\signalValue{\agent}{\timePoint}} of model \EVModel with state of charge \windowValue{\agent}{x}{\windowIndex} at the beginning time $x$ of window \windowIndex is given as follows:

\begin{equation}\label{eq:chargeRate}
\slotSize{\agent}{\windowIndex}=\chargeTime{\EVModel}{\windowValue{\agent}{x}{\windowIndex}}=(1-\windowValue{\agent}{x}{\windowIndex})\frac{\batteryCapacity{\EVModel}}{\chargeRate{\EVModel}},
\end{equation}

\noindent where \batteryCapacity{\EVModel} is the battery capacity of model \EVModel (kWh) and \chargeRate{\EVModel} is the charging rate of model \EVModel (kW). 

The usual operation of an electric vehicle suggests that within a window corresponding to `parked at home' or `parked at work', the driver charges immediately the vehicle. This action has significant implications. The power consumption mainly occurs at the very beginning of the window instead throughout the window time. Given that these windows among citizens have high temporal similarity as they correspond to a regular behaviorial activity and mobility patterns, the aggregate energy consumption at the beginning of the windows is synchronized among the electric vehicles and results in power peaks that can potentially cause blackouts or increase the energy cost. Moreover, as the battery technology~\cite{shen2014optimization} improves by allowing higher charging rates, the power peaks are expected to become even sharper. This section introduces a model that tackles this limitation by establishing \numOfWindowSlots{\agent}{\windowIndex} charging \emph{slots} $\slots{\agent}{\windowIndex}=(\slot{\agent}{\windowIndex}{\slotIndex})_{\slotIndex=1}^{\numOfWindowSlots{\agent}{\windowIndex}}$ within each window \windowIndex. The number of slots $\numOfWindowSlots{\agent}{\windowIndex}>0$ is computed by Algorithm~\ref{alg:slots-number}. Lines 2-6 determine the number of slots as follows:

\begin{equation}
\numOfWindowSlots{\agent}{\windowIndex}=\begin{cases}
|\window{\agent}{\windowIndex}|/\chargeTime{\EVModel}{\windowValue{\agent}{x}{\windowIndex}}, &\text{if $\numOfWindowSlots{\agent}{\windowIndex}<\numOfPlans$}\\
\numOfPlans, &\text{if $\numOfWindowSlots{\agent}{\windowIndex}\geq\numOfPlans$}
\end{cases},
\end{equation}

\begin{algorithm}[!htb]
\centering
\small{
\begin{algorithmic}[1]
\REQUIRE \windows{\agent}, \chargeTime{\EVModel}{\windowValue{\agent}{x}{\windowIndex}}, \numOfPlans
\FORALL{$\window{\agent}{\windowIndex} \in \windows{\agent}$}
\STATE $\numOfWindowSlots{\agent}{\windowIndex}=|\window{\agent}{\windowIndex}|/\chargeTime{\EVModel}{\windowValue{\agent}{x}{\windowIndex}}$
\IF{$\numOfWindowSlots{\agent}{\windowIndex} \geq \numOfPlans$}
\STATE \numOfWindowSlots{\agent}{\windowIndex}=\numOfPlans
\STATE $\slotSize{\agent}{\windowIndex}=|\window{\agent}{\windowIndex}|/\numOfWindowSlots{\agent}{\windowIndex}$
\ENDIF
\STATE $\numOfSlots{\agent}=\numOfSlots{\agent} \cup \numOfWindowSlots{\agent}{\windowIndex}$
\ENDFOR
\ENSURE \numOfSlots{\agent}, $\forall \agent \in \{1,...,\numOfAgents\}$ 
\end{algorithmic}
}
\caption{Computation of the number of slots for each window.}\label{alg:slots-number} 
\end{algorithm}

\noindent where $|\window{\agent}{\windowIndex}|$ is the size of window \windowIndex, \chargeTime{\EVModel}{\windowValue{\agent}{x}{\windowIndex}} is the charging time of model \EVModel at window \windowIndex and \numOfPlans determines the maximum number of plans with which an agent can operate. When knowing the number of slots per window, the actual slots can be determined according to Algorithm~\ref{alg:slots}. 


\begin{algorithm}[!htb]
\centering
\small{
\begin{algorithmic}[1]
\REQUIRE \windows{\agent}, \numOfSlots{\agent}, \slotSize{\agent}{\windowIndex}
\FOR{$\windowIndex=1$ to $\windowIndex=|\windows{\agent}|$}
\FOR{$\slotIndex=1$ to $\slotIndex=\numOfWindowSlots{\agent}{\windowIndex}$}
\FOR{$\timePoint=x+\slotIndex\cdot\slotSize{\agent}{\windowIndex}-\slotSize{\agent}{\windowIndex}$ to $x+\slotIndex\cdot\slotSize{\agent}{\windowIndex}-1$}
\STATE $\slot{\agent}{\windowIndex}{\slotIndex}=\slot{\agent}{\windowIndex}{\slotIndex} \cup \slotValue{\agent}{\timePoint}{\windowIndex}{\slotIndex}$
\ENDFOR
\STATE $\slots{\agent}{\windowIndex}=\slots{\agent}{\windowIndex} \cup \slot{\agent}{\windowIndex}{\slotIndex}$
\ENDFOR
\ENDFOR
\ENSURE \slots{\agent}{\windowIndex}, $\forall \agent \in \{1,...,\numOfAgents\}$ and $\forall \windowIndex \in \{1,...,\numOfWindows{\agent}\}$
\end{algorithmic}
}
\caption{Computation of slots.}\label{alg:slots} 
\end{algorithm}

Plan generation is then performed by redistributing the charging of electric vehicles among the slots. A design choice here is (i) how many slots to use and (ii) which slots to use. 

The number of slots determines the extent of redistribution, in other words, how uniformly distributed charging is over time. The plans are generated such that the first plan uses one slot, whereas the last plan uses all \numOfPlans slots. Each plan in between uses an additional slot incrementally. 

The slots used in each plan are determined by the likelihood of the vehicle usage $\EVUsage{\agent}=(\EVUsageValue{\agent}{\timePoint})_{\timePoint=1}^{\horizon}$ extracted from historical data. Based on these data, the slots in each window can be ranked from low to high likelihood of usage. Each plan uses the slots with the lowest likelihood utilization so that the likelihood of discomfort is minimized. Therefore, the discomfort \discomfort{\agent} of a plan can be defined as follows:

\begin{equation}\label{eq:discomfort}
\discomfort{\agent}=\frac{1}{\horizon}\sum_{\timePoint=1}^{\horizon}(1-\signalValue{\agent}{\timePoint})\cdot\EVUsageValue{\agent}{\timePoint},
\end{equation}

\noindent where \signalValue{\agent}{\timePoint} is the state of charge and \EVUsageValue{\agent}{\timePoint} is the likelihood of vehicle usage at time \timePoint. 

After determining the slots for each plan, charging in each window \windowIndex is performed in \chargeIntervals{\EVModel}{\windowValue{\agent}{x}{\windowIndex}} non-overlapping intervals of size \minChargingTime as follows:

\begin{equation}\label{eq:chargeIntervals}
\chargeIntervals{\EVModel}{\windowValue{\agent}{x}{\windowIndex}}=\frac{\chargeTime{\EVModel}{\windowValue{\agent}{x}{\windowIndex}}}{\minChargingTime},
\end{equation}

\noindent where \chargeTime{\EVModel}{\windowValue{\agent}{x}{\windowIndex}} is the charging duration and \minChargingTime is the minimum time interval that a vehicle is charged without pause. The intervals are uniformly distributed across the slots used by each plan. Algorithm~\ref{alg:plan-generation} illustrates the plan generation process. 

\begin{algorithm}[!htb]
\centering
\small{
\begin{algorithmic}[1]
\REQUIRE \signal{\agent}, \numOfSlots{\agent}, \slots{\agent}{\windowIndex}, \EVUsage{\agent}, \slotSize{\agent}{\windowIndex}
\FORALL{$\window{\agent}{\windowIndex} \in \windows{\agent}$}
\STATE \rankedSlots{\agent}{\windowIndex}=rank(\slots{\agent}{\windowIndex},\EVUsage{\agent})
\FOR{$\plan=1$ to $\plan=max(\numOfSlots{\agent})$ }
\STATE \shuffledSignal{\agent}=shuffle(\signal{\agent}, \plan, \rankedSlots{\agent}{\windowIndex}, \slotSize{\agent}{\windowIndex}, \minChargingTime)
\STATE $\demandPlan{\agent}{\plan}=\demandPlan{\agent}{\plan} \cup \demand{\agent}{\plan}{\timePoint}$
\ENDFOR
\ENDFOR
\ENSURE \demandPlans{\agent}, $\forall \agent \in \{1,...,\numOfAgents\}$ 
\end{algorithmic}
}
\caption{Computation of agent's plans.}\label{alg:plan-generation} 
\end{algorithm}

The plan generation algorithm iterates over the windows (line 1 of Algorithm~\ref{alg:plan-generation}) and ranks the slots of each window according to the likelihood \EVUsage{\agent} of vehicle usage (line 2 of Algorithm~\ref{alg:plan-generation}). The window with the maximum number of slots corresponds to the number of plans $\numOfPlans=max(\numOfSlots{\agent})$ (line 3 of Algorithm~\ref{alg:plan-generation}). Each plan \plan is generated by randomly shuffling \chargeIntervals{\EVModel}{\windowValue{\agent}{x}{\windowIndex}} charging intervals over \plan slots (line 4 of Algorithm~\ref{alg:plan-generation}). The plan is computed as $\demandPlan{\agent}{\plan}=\demandPlan{\agent}{\plan} \cup \demand{\agent}{\plan}{\timePoint}$, where:

\begin{equation}\label{eq:plan}
\demand{\agent}{\plan}{\timePoint}=\begin{cases}
\power{\EVModel}>0, &\text{if $\shuffledSignalValue{\agent}{\timePoint}<\shuffledSignalValue{\agent}{\timePoint+1}$}\\
0, &\text{if $\shuffledSignalValue{\agent}{\timePoint}\geq\shuffledSignalValue{\agent}{\timePoint+1}$}
\end{cases}, \forall \timePoint \in \{1,...,\horizon-1\}.
\end{equation}

\noindent Therefore, each value $\demand{\agent}{\plan}{\timePoint}\geq0$ of a plan contains the power demand \power{\EVModel} of model \EVModel (kW) at time \timePoint for the respective change $\shuffledSignalValue{\agent}{\timePoint}$ to $\shuffledSignalValue{\agent}{\timePoint+1}$ in the state of charge. 

\section{Decentralized Learning and Decision-making}\label{sec:decision-making}

This paper focuses on two system-wide objectives for the charging optimization of electric vehicles: (i) \minDev and (ii) \minCost. The former aims at minimizing the standard deviation, $\sigma$, of the total demand as a measure of load uniformity, load balancing and peak-shaving that contribute to the \emph{reliability} of the Smart Grid. The latter aims at reducing the total energy cost by taking into account temporal energy prices.

Agents employ the I-EPOS system~\cite{Pilgerstorfer2017,Pournaras2018} as a fully decentralized and privacy-preserving learning mechanism for coordinating the charging of electric vehicles. I-EPOS has been studied earlier in load-balancing of bike sharing stations~\cite{Pournaras2018} and in demand-response of residential energy consumption~\cite{Pournaras2014,pournaras2014decentralized,Pournaras2017}. In that Smart Grid scenario, the agents control individual home appliances or the aggregate demand of the household. In contrast, this paper contributes a new application of I-EPOS in Smart Grids and provides fundamental insights on how the charging of electric vehicles can be modeled as a 0-1 multiple-choice combinatorial optimization problem. In such a model, the optimization turns to be a plan selection problem of complexity $O(\numOfPlans^{\numOfAgents})$: each agents selects one of the possible charging plans such that the total consumption of all electric vehicles satisfy one of the two aforementioned objectives.

To manage this vast combinatiorial complexity, the agents of I-EPOS are self-organized~\cite{Pournaras2013b} in a tree topology as a way to structure their interactions with which they perform a cooperative optimization. A tree topology is a design choice to perform a computationally cost-effective aggregation of the power demand level as well as to perform coordinated decision-making. The computational and communication complexity depends on the number and size of plans \numOfPlans as well as the number of children \numOfChildren that each node has such that $O(\numOfPlans^{\numOfChildren})$, while the network topology does not have a significant impact on the performance as earlier shown~\cite{Pournaras2018}. This makes I-EPOS a highly efficient and scalable distributed algorithm for problems of combinatorial complexity as confirmed with comparisons to other state of the art algorithms~\cite{Pournaras2018}.

The optimization of the plan selections is performed by a set of consecutive \emph{learning iterations} of \emph{bottom-up} (leaves to root) and \emph{top-down} (root to leaves) interactions. At each iteration, an agent \agent selects a plan $j$ to satisfy the \minDev optimization objective as follows:

\begin{equation}\label{eq:min-dev}
j=\argmin_{\plan=1}^{\numOfPlans}\sigma(\oldAggregatePlan{1}-\oldAggregatePlan{\agent}+\aggregatePlan{\agent}-\oldDemandPlan{\agent}+\demandPlan{\agent}{\plan})
\end{equation}

\noindent where $\sigma()$ measures the standard deviation of a plan, $\oldAggregatePlan{1}=\sum_{i=1}^{\numOfAgents}\oldDemandPlan{\agent}$ is the earlier \emph{aggregate plan} of all earlier selected plans $\oldDemandPlan{\agent}$ summed up at the root $\agent=1$, $\oldAggregatePlan{\agent}$, $\aggregatePlan{\agent}$ are the earlier and current aggregate plans respectively of all plan selections of the agents in the branch underneath agent \agent and \oldDemandPlan{\agent}, \demandPlan{\agent}{\plan} are the earlier selected plan and the current possible plan \plan of agent \agent. Note that the minimization of variance and standard deviation are \emph{quadratic cost functions}~\cite{Rockafellar2000} that requires coordination among the agents' selections. The aggregate plans in Equation~\ref{eq:min-dev} serve this purpose. Moreover, privacy is preserved by only exchanging aggregate plans instead of the individual ones. Further elaboration on the I-EPOS algorithm is out of the scope of this paper and is available on earlier work~\cite{Pilgerstorfer2017}. 

The \minCost selection function aims at reducing the total energy cost by taking into account the temporal energy prices as follows:

\begin{equation}\label{eq:min-cost}
\plan=\argmin_{\plan=1}^{\numOfPlans}\sum_{\timePoint=1}^{\horizon}(\aggregateDemand{\agent}{\timePoint}+\demand{\agent}{\plan}{\timePoint})\cdot \priceValue{\timePoint},
\end{equation}

\noindent where \aggregateDemand{\agent}{\timePoint} is the power demand of the aggregate plan at time \timePoint, \demand{\agent}{\plan}{\timePoint} is the power demand of the possible plan \plan at time point \timePoint and \priceValue{\timePoint} is the energy price at time \timePoint. Note that this is a \emph{linear cost function} that can be minimized locally without requiring coordination among agents, i.e. the minimum total energy cost computed at the end of the first learning iteration is optimal and therefore no further iterations are required. The use of I-EPOS in this case serves exclusively the distributed aggregation of the selected plans \aggregatePlan{\agent} and therefore, the term \aggregateDemand{\agent}{\timePoint} is not actually required for the optimization.


This paper studies how the optimization of reliability using the \minDev and \minCost objectives may influence human and social aspects such as the discomfort and fairness respectively. The \emph{system discomfort} \totalDiscomfort is measured by the average discomfort as follows:

\begin{equation}\label{eq:total-discomfort}
\totalDiscomfort=\frac{1}{\numOfAgents}\sum_{\agent=1}^{\numOfAgents}\discomfort{\agent}.
\end{equation}

\noindent A set of charging regimes are defined as fair if all agents have the same level of discomfort. Fairness increases with the reduction in the dispersion of discomfort. Mathematically, fairness is defined as follows:

\begin{equation}\label{eq:fairness}
\fairness=1-\sigma(\discomfort{1},...,\discomfort{\numOfAgents}),
\end{equation}

\noindent where $\sigma(\discomfort{1},...,\discomfort{\numOfAgents})$ measures the standard deviation of the discomfort values among the agents. 

Note that other more complex objective functions for reliability could be employed such as scenarios of power generator failures~\cite{Pournaras2017} and cascading failures triggered by power line failures~\cite{Thapa2017}. In such scenarios I-EPOS can optimize a matching objective between power demand and a given incentive signal computed by parametric power supply models of transactive control systems~\cite{Pournaras2017}. Such models are made available within the I-EPOS software artifact~\cite{Pournaras2018}.

\section{Results and Discussion}\label{sec:evaluation}

This section illustrates the experimental methodology based on a real-world dataset and the experimental results of the decentralized optimization. 

\subsection{Experimental methodology}\label{subsec:methodology}

\begin{figure}[!htb]
\centering
\includegraphics[width=2.5in]{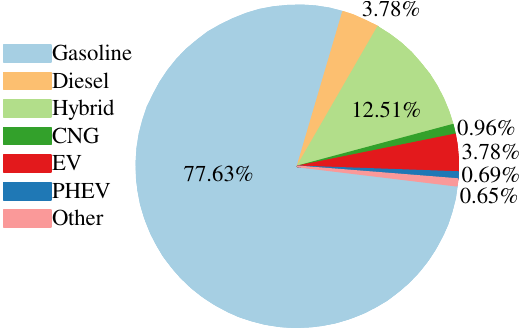}
\caption{Vehicles by type contained in the California Household Travel Survey for 2010–-2012~\cite{NRELSurvey2015}. EV: Electric Vehicle, PHEV: Plug-in Hybrid Electric Vehicle, CNG: Compressed Natural Gas.}
\label{fig: vehicle_breakdown}
\end{figure}

The conducted experiments\footnote{The raw data of all experimental results illustrarted in the plots of this paper as available in the Supplementary Material.} are based on real-world data from the California Household Travel Survey of the California Department of Transportation during the period 2010--2012~\cite{NRELSurvey2015}. The data cover several aspects of electric vehicles, citizens' demographics, mobility, information about the vehicles and other. The vehicle types are outlined in Figure~\ref{fig: vehicle_breakdown}. A number of 2910 vehicles are fitted with GPS with a recording resolution of one second: This monitors vehicles continuously for a period of seven days, although the days may be different among the vehicles. 

\begin{figure}[!htb]
\centering
\subfigure[Vehicle usage]{\includegraphics[width=0.49\columnwidth]{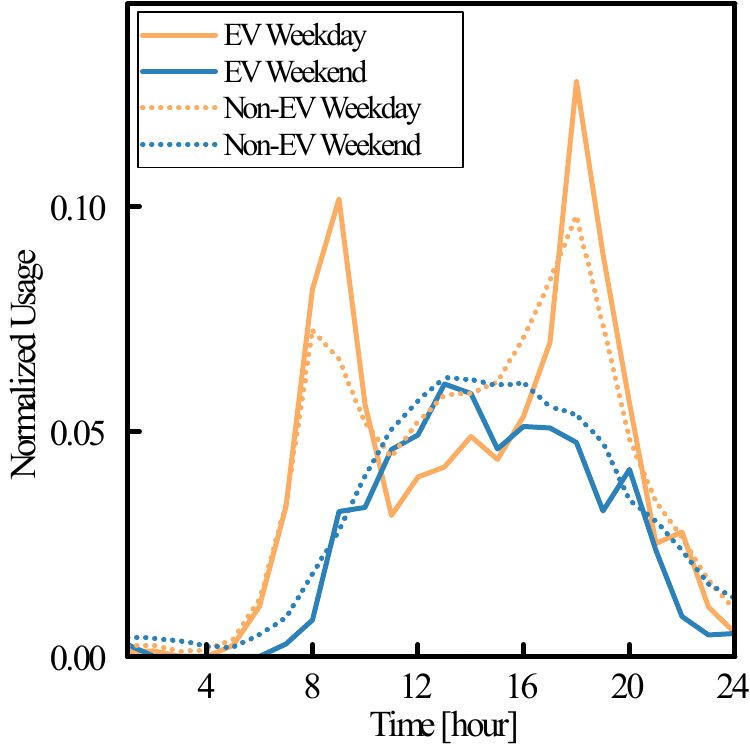}}
\subfigure[Driver profile]{\includegraphics[width=0.49\columnwidth]{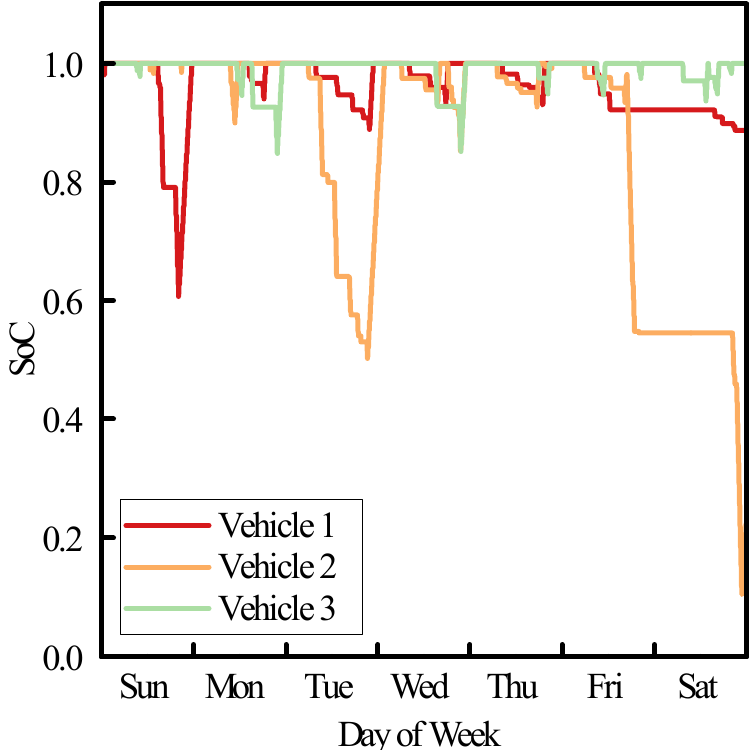}}
\caption{(a) The use of electric vehicles throughout the week. For comparison, the usage of non-electric vehicles is also shown. Data for all vehicles are binned within one minute periods. The usage frequency at each minute is divided by the total number of vehicles used. Note the characteristic peaks on rush hours during the week, the usage during the day and the intuitive daytime usage during the weekend days. The qualitative shape of the curves remains the same among the vehicle types i.e. electric vs. non-electric. Both electric and plug-in hybrid vehicles are under the `EV' category. (b) Three state of charge (SoC) profiles over a period of one week for several vehicles of the same vehicle model: Tesla S~\cite{TeslaPower}.}\label{fig: usage-soc}
\end{figure}

The GPS data are preprocessed to compute the trip profiles of each vehicle. The profiles encode each trip made by the vehicle as a \emph{destination} i.e. home, work, school, or other, a \emph{start/finish} trip time, and an \emph{average speed}. From these data, the usage of vehicles can be analyzed on a weekly basis given their type. Weekly\footnote{The model of Section~\ref{sec:model-concept} relies on historical driver data to calculate the probabilistic availability of each vehicle throughout the week. A limitation of the Californian Survey dataset is that the data per vehicle is only a week in length and sampled at different weeks. Therefore, it is assumed that the data are representative of a weekly usage of the vehicle. However, this limitation does not influence the design of operational planning or the decentralized optimization approach that are data independent. Future work can extend the findings of this paper with further new datasets~\cite{Xu2018,Kujala2018}.} usage of all GPS-equipped vehicles in the dataset are shown in Figure \ref{fig: usage-soc}a.

Given the trip profiles, the state of charge profiles can be constructed for the entire week from Sunday 00:00 to Saturday 23:59 with the following assumptions: (i) Every vehicle starts at 100\% state of charge on Sunday at 00:00. (ii) Charging starts as soon as a vehicle arrives at home. (iii) Every vehicle of the same type is assumed to have the same technical specifications, for example, battery charge/discharge rates, battery capacity, etc, as listed in Table~\ref{table: EVmodels}. (iv) A vehicle in transit travels at the average speed during the entire journey as reported in the dataset. (v) A vehicle faster than 60 mph discharges according to its highway mileage, otherwise according to its city mileage. The `miles per gallon' denotes the efficiency of the electric vehicle as outlined in Table~\ref{table: EVmodels}. From the original pool of 2910 GPS-equipped vehicles, 131 vehicles are excluded that contain a negative state of charge value no matter which vehicle model is assigned to them. The remaining 2779 vehicles are assigned to one of the five models listed in Table~\ref{table: EVmodels}. The shares of electric vehicle models in this final pool follows the current market shares of the vehicle models. 

\begin{table}[!htb]
\caption{Descriptive statistics of the electric vehicle models. MPG: Miles Per Gallon of gasoline, \batteryCapacity{\EVModel}: battery capacity and \chargeRate{\EVModel}: charge rate.}\label{table: EVmodels}
{\footnotesize
\centering
\begin{tabular}{|m{2.5cm}|m{2.0cm}|m{1.3cm}|m{1.3cm}|}
\hline
Model & MPG City/Highway & \batteryCapacity{\EVModel} (kWh) & \chargeRate{\EVModel} (kW) \\ \hline
\hline
Nissan Leaf~\cite{NissanPower} & 126/101 & 24 & 6.6 \\
Tesla S 85 kWh \cite{TeslaPower}& 88/90 & 85 & 9.6\\
BMW i3~\cite{BMWPower}& 137/111 & 22 & 7.4 \\
Fiat 500e~\cite{FiatPower} & 121/103 & 24 & 6.6 \\
Ford Focus~\cite{FordPower} & 110/99 & 23 & 6.6 \\
\hline
\end{tabular}
}
\end{table}

The energy consumption (in kWh) of a trip is given as follows:

\begin{equation}\label{eq: trip-energy}
E_\mathtt{t} = \frac{d \cdot \eta}{f_e},
\end{equation}

\noindent where $\eta = 33.705$ kWh/gallon (the conversion of energy between gasoline and Joules~\cite{DOE_formula}), $d=s\cdot t_\mathtt{t}$ is the distance covered during the journey, with $s$ the average speed in mph and $t_\mathtt{t}$ the duration of the journey in hours. The $f_e$ (miles/gallon) is the fuel efficiency of the vehicle model for a given scenario: either moving in the city or on a highway (see Table~\ref{table: EVmodels}). Given that the vehicular trip data contain an average speed, $f_e$ is determined as moving in the city when the average speed is $s\leq 60$ mph or moving on the highway if $s > 60$mph. Moreover, given the energy consumption of every trip in Equation~\ref{eq: trip-energy}, the state of charge of the battery can be calculated as a function of time based on the battery capacity and battery charging rate of a vehicle (Table~\ref{table: EVmodels}). An illustration of three profiles about the state of charge of a vehicle are given in Figure~\ref{fig: usage-soc}b.

The aforementioned preprocessed data are used to generate the plans as shown in Section~\ref{sec:model}. The plans are made openly available~\cite{Pournaras2019b} for the community to further study the charging control of electric vehicles as well as new algorithms for combinatorial optimization problems. I-EPOS\footnote{Available at \url {http://epos-net.org} and \url{https://github.com/epournaras/EPOS} (last accessed: January 2019)}~\cite{Pournaras2019c} is implemented in the Protopeer distributed prototyping toolkit~\cite{Galuba2009}. The agents are randomly positioned in a binary tree topology. All agents generate four plans\footnote{A sensitivity analysis for the tree topology and the number of plans is carried out in earlier work~\cite{Pournaras2018} confirming in various settings that (i) the topological structure plays no significant role in the optimization and (ii) a higher number of possible plans provides a higher degree of freedom for the further reduction of the cost function.}. Four participation scenarios are evaluated. Each scenario assumes that a subset of all electric vehicles is equipped with the capability to generate plans. The rest of the non-participating vehicles use the default charging pattern observed in the historical data. The four participation levels are 25\%, 50\%, 75\% and 100\%. The planning horizon is set to $\horizon=1440$ and $\horizon=10080$ that correspond to daily\footnote{Under the daily optimization scheme, the week is split into 8 time periods: Sunday 00:00-11:59 (12 hours), Sunday 12:00-Monday 11:59 (24 hours), Monday 12:00-Tuesday 11:59 (24 hours), Tuesday 12:00-Wednesday 11:59 (24 hours), Wednesday 12:00-Thursday 11:59 (24 hours), Thursday 12:00-Friday 11:59 (24 hours), Friday 12:00-Saturday 11:59 (24 hours) and Saturday 12:00-23:59 (12 hours). } and weekly optimization respectively. 

\subsection*{Experimental results}

\begin{figure}[!htb]
\centering
\includegraphics[width=0.82\columnwidth]{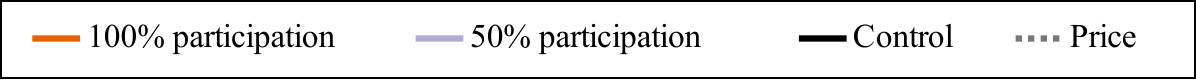}\\
\subfigure[\minDev, daily]{\includegraphics[width=0.49\columnwidth]{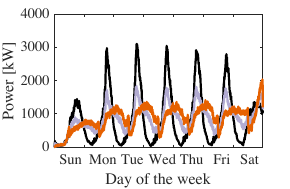}}
\subfigure[\minCost, daily]{\includegraphics[width=0.49\columnwidth]{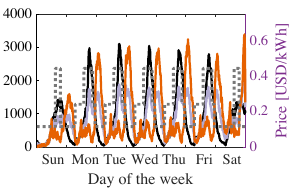}}
\subfigure[\minDev, weekly]{\includegraphics[width=0.49\columnwidth]{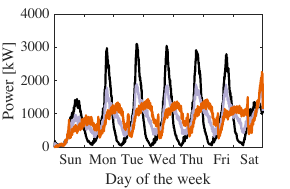}}
\subfigure[\minCost, weekly]{\includegraphics[width=0.49\columnwidth]{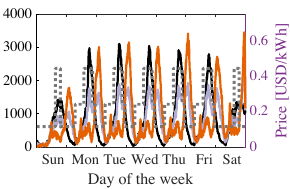}}
\caption{Demand curves of the electric vehicles under daily and weekly optimization with \minDev and \minCost.}\label{fig:curves}
\end{figure}

\begin{figure}[!htb]
\centering
\subfigure[\minDev]{\includegraphics[width=1.0\columnwidth]{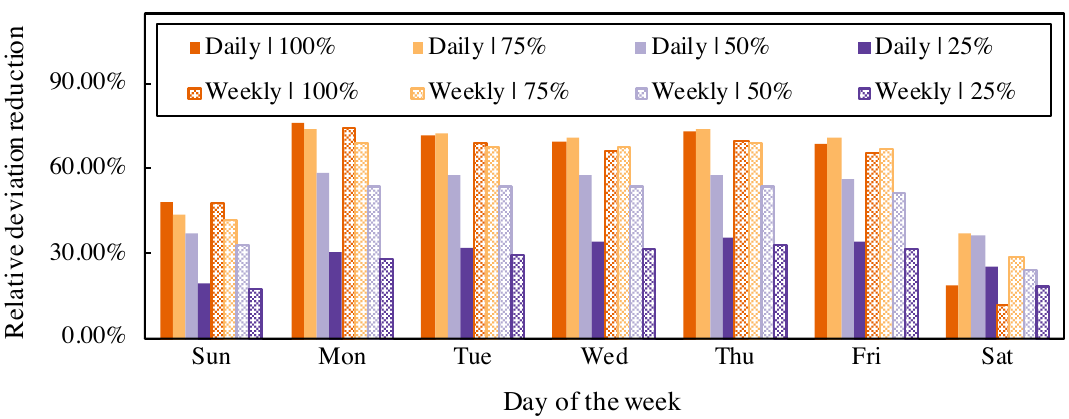}}
\subfigure[\minCost]{\includegraphics[width=1.0\columnwidth]{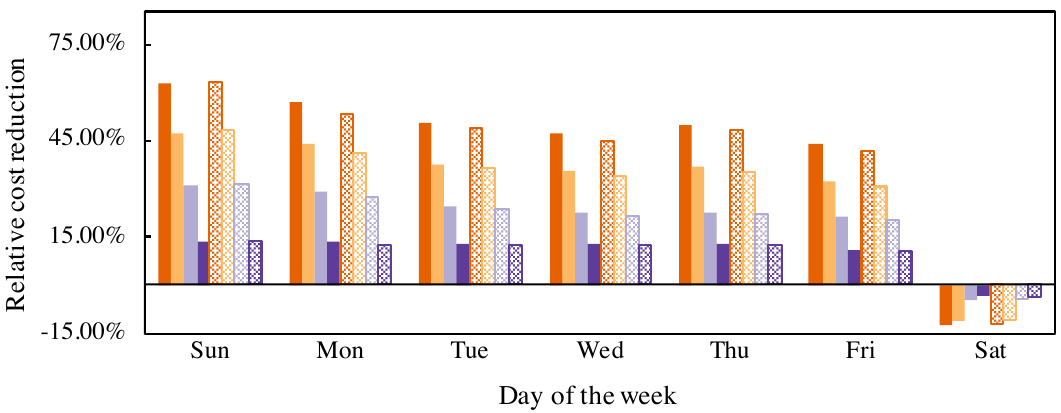}}
\caption{Performance of I-EPOS (10th interation) under a varied participation level.}\label{fig:performance}
\end{figure}

An overview of the load curves is given in Figure~\ref{fig:curves}. Figure~\ref{fig:performance} illustrates the performance of I-EPOS under \minDev (10th iteration) and \minCost. The performance is measured by the relative, to the control data, decrease in standard deviation and cost respectively. The relative deviation reduction under daily optimization is on average 30.1\%, 51.7\%, 63.3\% and 61.1\% for 25\%, 50\%, 75\% and 100\% participation level. Respectively, the relative deviation reduction under weekly optimization is 27.1\%, 46.1\%, 58.7\% and 57.7\%. The relative cost reduction under daily optimization is on average 10.5\%, 21.1\%, 31.8\% and 42.7\% for 25\%, 50\%, 75\% and 100\% participation level. For weekly optimization, the respective cost reduction is 10.1\%, 20.4\%, 30.7\% and 41.2\%. For the \minDev and \minCost optimization goals, the performance of daily optimization is on average 3.54\% and 8.74\% higher respectively than the weekly one over all participation levels. 


Note that the negative cost reduction on Saturday (Figure~\ref{fig:performance}b) is an artifact of the limited data: The agents avoiding the peak spot price on Friday choose to charge on Saturday morning given the off-peak price. However, the load on Saturday cannot be shifted to the off-peak hours on Sunday morning as Saturday is the last day in the performed experiments. The global power demand curves in Figure~\ref{fig:curves} confirm this rationale. In case the performed experiments were not bound to one week and the Saturday load could be shifted to the next morning--a more realistic scenario--the cost reduction on Saturday should be positive as well.

\begin{figure}[!htb]
\centering
\includegraphics[width=1.0\columnwidth]{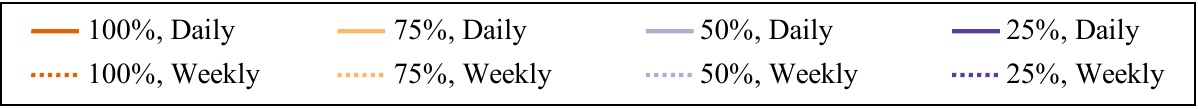}
\subfigure[\minDev, relative deviation]{\includegraphics[width=0.49\columnwidth]{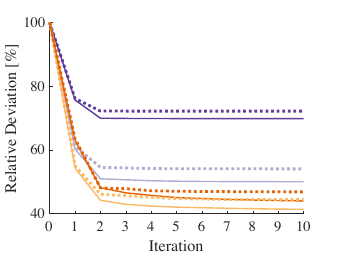}}
\subfigure[\minCost, relative cost]{\includegraphics[width=0.49\columnwidth]{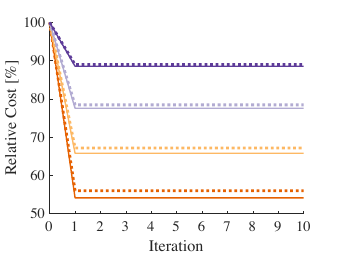}}
\subfigure[\minDev, discomfort]{\includegraphics[width=0.49\columnwidth]{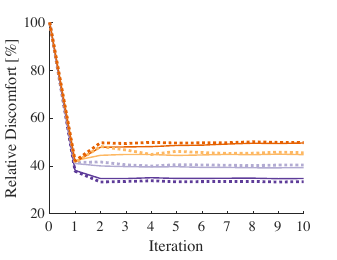}}
\subfigure[\minCost, discomfort]{\includegraphics[width=0.49\columnwidth]{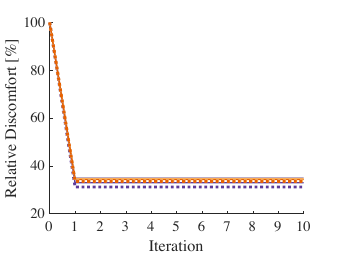}}
\subfigure[\minDev, fairness]{\includegraphics[width=0.49\columnwidth]{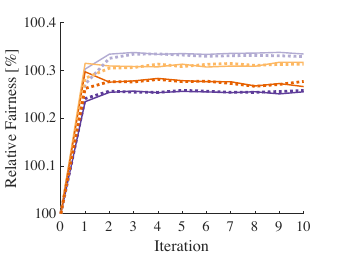}}
\subfigure[\minCost, fairness]{\includegraphics[width=0.49\columnwidth]{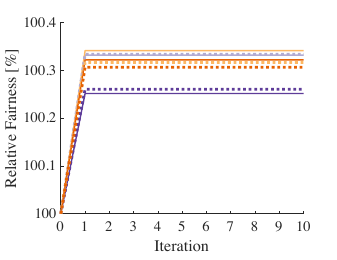}}
\caption{Learning curves for different levels of participation.}\label{fig:learning-curves}
\end{figure}

Figure~\ref{fig:learning-curves}a,~\ref{fig:learning-curves}c and~\ref{fig:learning-curves}e illustrate the learning curves in terms of relative deviation, discomfort and fairness respectively for \minDev under varied levels of participation. Note that optimization is performed to improve the relative deviation. The influence of discomfort and fairness over the course of the iteration is illustrated. Moreover, note that although learning is not performed under the linear cost function of \minCost, results are included in Figure~\ref{fig:learning-curves}b,~\ref{fig:learning-curves}d and~\ref{fig:learning-curves}f for more comprehensive comparisons. 

Figure~\ref{fig:learning-curves}a confirms the monotonous improvement of reliability by \minDev in the course of 10 iterations. Convergence though is performed rapidly in the first 3-4 iterations. This learning trend is similar for varied participation levels. Discomfort and fairness in Figure~\ref{fig:learning-curves}c and~\ref{fig:learning-curves}e remain stable over the course of learning iterations.

\begin{figure*}[!htb]
\centering
\hfill\includegraphics[width=0.91\textwidth]{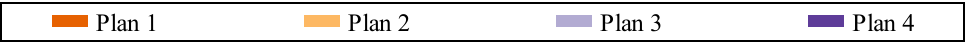}
\subfigure[\minDev]{\includegraphics[width=1.0\textwidth]{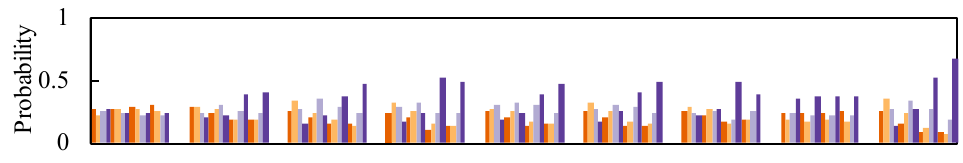}}
\subfigure[\minCost]{\includegraphics[width=1.0\textwidth]{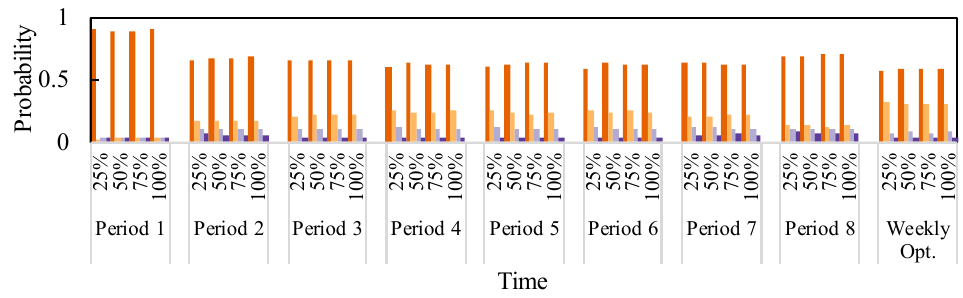}}
\caption{Plan selections under varied participation level.}\label{fig:selections}
\end{figure*}

Figure~\ref{fig:selections} illustrates the probability of plan selections in the performed experiments. In \minDev with a 100\% participation level, Plans 1, 2, 3, and 4 are selected with 0.18, 0.16, 0.23, and 0.43 probability under daily optimization and 0.08, 0.07, 0.18, and 0.66 probability under weekly optimization. In contrast, the respective probabilities for \minCost change as follows: 0.66, 0.20, 0.10, and 0.04 under daily optimization and 0.58, 0.31, 0.08, 0.04 under weekly optimization. Under \minDev, Plan 4 is the most frequently selected plan as it gradually charges vehicles over all available charging slots. On the other hand, Plan 1 is the most frequently selected one under \minCost as a single charging slot makes more likely the completion of charging during off-peak hours. 

\begin{figure*}[!htb]
\centering
\hfill\includegraphics[width=0.91\textwidth]{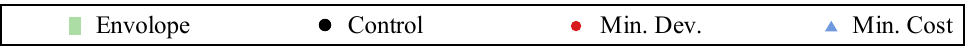}
\includegraphics[width=1.0\textwidth]{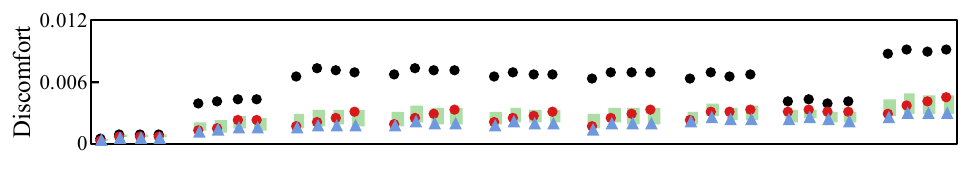}
\includegraphics[width=1.0\textwidth]{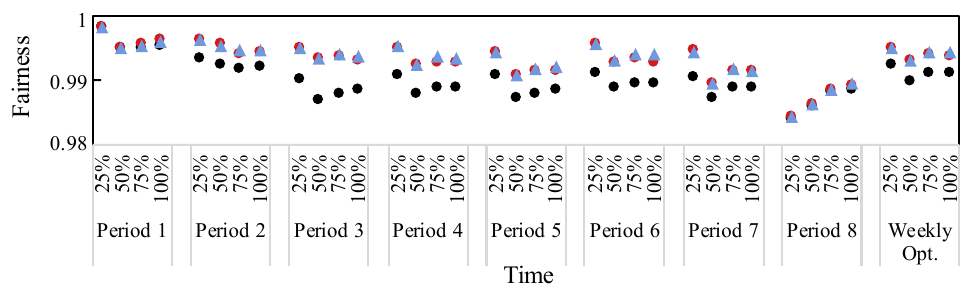}
\caption{Discomfort and fairness of \minDev, \minCost and control data under varied participation level. The discomfort envelope defines the upper (Plan 4) and lower (Plan 1) bounds.}\label{fig:discomfort-fairness}
\end{figure*}

Figure~\ref{fig:discomfort-fairness}a illustrates the mean discomfort for \minDev, \minCost and control data under a varying participation level. A discomfort envelope is defined by the upper and lower bounds when all agents select Plan 4 and Plan 1 respectively. This is because the slots used for charging are ranked according to the likelihood of utilization in the historical data. 

Given that Plan 1 is frequently selected under \minCost and Plan 4 under \minDev (Figure~\ref{fig:selections}), the mean discomfort of \minCost is closer to the lower bound and \minDev closer to the upper bound of each envelope. \minCost has on average 19.0\% lower discomfort than \minDev. Moreover, the striking lower discomfort of I-EPOS compared to controlled data is a result of the plan generation design and the ranking of the slots as well. By taking a careful look at Figure~\ref{fig:generation}, it can be observed that Figure~\ref{fig:generation}b in Window 3 has higher discomfort than the plan of Figure~\ref{fig:generation}c with zero discomfort because of the likelihood of utilization during charging. Moreover, the mean discomfort for \minCost is on average 0.00168, 0.00204, 0.00197 and 0.00197 for 25\%, 50\%, 75\% and 100\% participation level. A similar trend is confirmed for \minDev. This means that a low number of participating agents has to make more disruptive decisions to anticipate for the missing contributions of the non-participating agents. The shift of the selection from Plan 4 under 100\% participation level towards Plan 1 and 2 further justifies this finding.

Note that the low scale of the discomfort values in Figure~\ref{fig:discomfort-fairness}a are a result of how discomfort is defined in Equation~\ref{eq:discomfort}, i.e. a multiplication of the state of charge with the likelihood of the vehicle utilization by the selected plan, given that both of these measures vary in the range $[0,1]$. As an illustrative example: For plans charging at slots with 0.1 likelihood of vehicle utilization, the range $[0,0.012]$ of discomfort values in Figure~\ref{fig:discomfort-fairness}a corresponds to vehicles charged up to 88\%, which is likely to cause a range anxiety to some drivers~\cite{Neubauer2014,Rauh2015,Jung2015}. This information can be used in multi-dimensional incentive and cryptocurrency schemes~\cite{Munsing2017} designed for both collective and individual participation criteria: a citizen may have an intrinsic interest in Smart Grid reliability while a higher battery charging level can be achieved via the optimization.

Figure~\ref{fig:discomfort-fairness}b illustrates the fairness for \minDev, \minCost and control data under varied participation levels. Fairness improves for both \minDev and \minCost compared to the control data. \minCost has on average 0.038\% higher fairness than \minDev. The mean fairness for \minCost is on average 0.9960, 0.9966, 0.9962 and 0.9960 for 25\%, 50\%, 75\% and 100\% participation level. A similar trend is confirmed for \minDev.

%

\section{Conclusion and Future Work}\label{sec:conclusion}

This paper concludes that a socio-technical Smart Grid optimization is feasible by decentralized coordination of charging control in electric vehicles. This paper introduces a novel planning mechanism for battery charging of electric vehicles running locally by an autonomous software agent. Collective decision-making of the executed plans by the decentralized learning system of I-EPOS achieves significantly lower power peaks and energy costs without violating citizens' privacy. Social welfare aspects such as discomfort and fairness are measured and regulated making the applicability of the proposed methodology more realistic in complex socio-technical system such as the Smart Grid.

Social welfare in terms of discomfort and fairness improve in all scenarios compared to the empirical observations in real-world data. The experimental findings show that by minimizing the power peaks a higher discomfort is introduced for citizens compared to minimizing power costs. This is because making the power load more uniformly distributed requires the charging of vehicles at times of a higher likelihood of usage. The experimental findings also confirm and quantify the potential for higher reliability and cost reduction when a larger number of electric vehicles participate in the optimization process. These findings are relevant for power utilities, system operators and technology stakeholders in the business ecosystem of Smart Grids and the broader application domain of electric vehicles.

Future work aspires to study the socio-technical optimization capacity of I-EPOS in more complete datasets~\cite{Xu2018,Kujala2018} as well as in pilot field tests. The applicability of more complex parametric models for reliability is subject of future work. The integration of decentralized charging control of electric vehicles with residential demand-side energy management applications and the utilization of renewable energy resources can further expand the operational flexibility of the Smart Grid. In addition, charging regimes may play a key role on vehicle mobility patterns as well as traffic jams in future traffic flow systems. The modeling of the charging coordination as a bottom-up sharing economy can bring new opportunities for active citizens' participation in self-sustained Smart Cities.

\section*{Acknowledgment}

The authors would like to thank Lloyd Sanders for all fruitful discussions and his overall support on this project. This work is supported by the European Community’s H2020 Program under the scheme ‘INFRAIA-1-2014-2015: Research Infrastructures’, grant agreement \#654024 ‘SoBigData: Social Mining \& Big Data Ecosystem’ (\url{http://www.sobigdata.eu}) and the European Community’s H2020 Program under the scheme ‘ICT-10-2015 RIA’, grant agreement \#688364 ‘ASSET: Instant Gratification for Collective Awareness and Sustainable Consumerism’ (\url{http://www.asset-consumerism.eu}).

\bibliography{ev-optimization}                                              %

\end{document}